# Second-order phase transitions, L. Landau and his successors

Yuri Mnyukh
*76 Peggy Lane, Farmington, CT, USA, e-mail: yuri@mnyukh.com*
(Dated: February 5, 2011)

There are only two ways for solid-state phase transitions to be compliant with thermodynamics: emerging of infinitesimal *quantity* of the new phase, or infinitesimal "*qualitative*" change occurring uniformly throughout the bulk at a time. The suggested theories of phase transitions are checked here for that compliance and in historical perspective. While introducing the theory of "continuous" *second-order* phase transitions, L. Landau claimed that they "may also exist" along with the majority of *first order* phase transitions, the latter being "discontinuous", displaying "jumps" of their physical properties; the fundamental differences between the two types were specified. But his theoretical successors disregarded these irreconcilable differences, incorrectly presenting all phase transitions as a cooperative phenomenon treatable by statistical mechanics. In the meantime, evidence has been mounted that *all* phase transitions have a nucleation-and-growth mechanism, thus making the above classification unneeded.

## 1. Compliance with thermodynamics

Physicists in the beginning of 20th century knew that phase transitions in solid state are not "continuous" in nature. But starting from 1930's the idea of "continuous" phase transitions emerged.

When contemplating possible mechanisms of phase transitions, it should be first realized that they have, as minimum, to meet the following conditions in order to comply with thermodynamics. An infinitesimal change of a controlling parameter (dT in case of temperature) may produce only two results: either (*A*) an infinitesimal *quantity* of the new phase emerges, with the structure and properties changed by finite values, or (*B*) a physically infinitesimal "*qualitative*" change occurs uniformly throughout the whole macroscopic bulk [1]. The conditions, however, do not guarantee both versions to exist in nature.

The version *'A'* is, evidently, an abstract description of the usually observed phase transitions by nucleation and growth. Every input of a minuscule quantity of heat $\delta Q$ either creates a nucleus or, if it exists, shifts the interface position by a minuscule length $\delta\ell$. The issue is whether version *'B'* can actually materialize. As far back as 1933, Ehrenfest formally classified phase transitions by *first-order* and *second-order* in terms of "continuity" or "discontinuity" in their certain thermodynamic functions [2]. It was a theoretical exercise; the validity of the classification was disputed by Justi and Laue by asserting that there is no thermodynamic or experimental justification for second-order phase transitions [3]. Judging from the absence of references in subsequent literature, their objections were ignored.

## 2. Second-order phase transitions: "may also exist"

Landau [4-6] developed a theory of *second-order* phase transitions. But he emphasized that transitions between different crystal modifications are "usually" *first-order*, occurring by sudden rearrangement of the crystal lattice at which the state of the matter changes abruptly, latent heat is absorbed or released, symmetries of the phases are not related and overheating or overcooling is possible. As for *second-order* phase transitions, they "may also exist", but no incontrovertible evidence of their existence was presented. It should be noted that expression that something "may exist" implicitly allows it not exist either. In case *second-order* phase transitions do exist, they must occur homogeneously, without any overheating or overcooling, at "critical points" where only the crystal symmetry changes, but structural change is infinitesimal. Landau left no doubt that his theory is that of *second-order* phase transitions only.

Since then it became accepted that there are "discontinuous" *first-order* phase transitions, exhibiting "jumps" in their physical properties, as well as "continuous" *second-order* phase transitions without "jumps". The latter are to be identified with the version *'B',* for they fit that particular version and, besides, no other option exists. Leaving alone the theory itself, there were several shortcomings in the Landau's presentation:

♦ He had not answered the arguments of the contemporaries, Max von Laue among them, that second-order phase transitions do not - and cannot - exist.

- The only examples he used to illustrate second-order phase transitions, $NH_4Cl$ and $BaTiO_3$, both turned out to be first order.
- The theory was unable to explain so called "heat capacity λ-anomalies" which, it should be noted, appeared also in first-order phase transitions.
- The description of first-order phase transitions left false impression that the "jump-like" changes occur simultaneously over the bulk.
- Overheating and overcooling in first-order transitions are not only "possible", they are inevitable (hysteresis).
- It was not specified that the only way first-order phase transitions can materialize is *nucleation and growth*;
- He remained silent when other theorists began to "further develop" his theory by treating the transitions of both types as a "critical phenomenon" in clear violation of the basic assumption of the classification in question.

.

### 3. First-order phase transitions in more detail

In order to better evaluate the ensuing chain of events, we need to expand Landau's characterization of first-order phase transitions by their features revealed in the subsequent studies [7-20] summarized in [21]. Solid-state phase transitions are realized by a crystal growth involving nucleation and propagation of interfaces. Nucleation is not the classical fluctuation-based process described in the textbooks. Nucleation in a given crystal is a pre-determined process. The nuclei are located in specific crystal defects - microcavities of a certain optimum size. These defects already contain information on the condition (*e.g.*, temperature) of their activation and on orientation of the resultant crystal lattice. Nucleation lags are inevitable and reproducible for a given defect, but are not the same in different defects. The transition is an intrinsically *local* process. It proceeds by "molecule-by-molecule" structural rearrangement at interfaces only, while the bulks of the original and emerged phases remain static (Fig.1). No macroscopic "jumps" occur during the phase transition. They are simply the differences between physical properties of the initial and resultant phases, revealing themselves as "jumps" when the transition range is narrow enough.

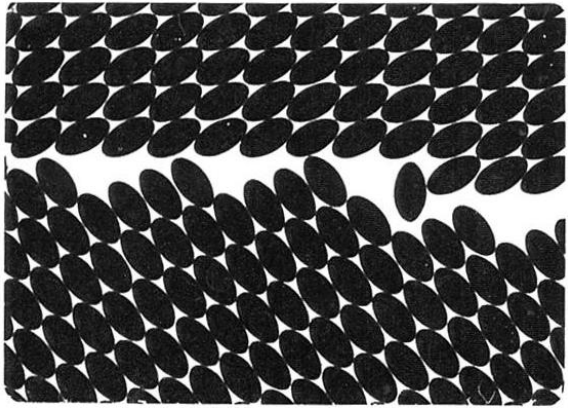

Fig. 1. Molecular model of phase transition in a crystal. The contact interface is a rational crystal plane in the resultant phase, but not necessarily in the initial phase. The interface advancement has the edgewise mechanism. It proceeds by shuttle-like strokes of small steps (kinks), filled by molecule-by-molecule, and then layer-by-layer in this manner. (Crystal growth from liquids is realized by the same mechanism). Besides the direct contact of the two different structures, existence of the 0.5 molecular layer gap (on average) should be noted. It is wide enough to provide steric freedom for the molecular relocation at the kink, but it is narrow enough for the relocation to occur under attraction from the resultant crystal. More detailed description of the process and its advantages is given in Ref. 21 (Sec. 2.4.2-2.4.6 ).

### 4. How to identify the 'order'

In order to distinguish between *first* and *second* order transitions, an indicator is needed capable to tell whether the process is local or homogeneous. The reliable indicators of first-order phase transitions are *interface, heterophase state*, and *hysteresis* - any one is sufficient, for all three are intimately linked. Thus, in principle, identification of a first-order transition is simple and definite. Not so with second-order transitions requiring proving that the above indicators are absent, whiling they can be overlooked or remain beyond the instrumental capability. The same is true for a property "jump". Its absence cannot serve as an indicator of second-order transition. Even though the participated phases are not related, the "jump" can still be tiny. The ferromagnetic phase transition in Fe at 769 °C is a good example. For decades it was regarded as the best representative of second-order phase transitions. But it was established in 2001 that it is a nucleation-and-growth phase transition, even though no "jumps" were ever reported [22]. Several years later a small latent heat - an undeniable attribute of a first-order phase transition - was recorded [23]. It is small or undetected "jumps" that were the source of erroneous classifications of phase transitions as being second-order. This method lacks the ability to tell whether the process has a local or homogeneous nature.

However small the jump is, or even looking zero, it is not an indicator of the phase transition order. Considering that a second-order phase transition is incompatible with a phase coexistence at any temperature, detection of a simultaneous presence of the two phases in any proportion at any temperature would proof the nucleation-and-growth mechanism.

Presently, it can be asserted with confidence that proper verification of the remaining "second- order" phase transitions will turn them to first order. A steady process of second-to-first-order reclassification is going on. No case of reclassification in the opposite direction is known.

### 5. Blurring the boundaries

The Landau theory initiated an avalanche of theoretical papers and books, presented not as a "theory of *second-order* phase transitions", but as a "theory of phase transitions". The first-order transitions were incorporated as a "critical phenomenon" as well. The restrictions clearly expressed by Landau that a theory of second-order transitions is not applicable to first-order ones were circumvented. Thus, Bruce and Cowley [24] avoided the "order" problem by simple replacement of the original Landau's heading [4,5] "Phase Transitions of the Second Kind" (*i. e*., second order) by the "Landau Theory" to apply it to all phase transitions. The same road was taken by J.C. and P. Toledano [25]. Statistical mechanics was applied to many first-order transitions on the grounds that they are "almost", or "nearly", or "close to", second-order. Or, as Buerger specified, they are "90% second-order and 10% first-order" [26]. Such statement as "Although the Landau theory assumes continuous second-order phase transitions, it can be applied to weakly first-order transitions" [27] was typical. Even the very book by Landau and Lifshitz [6] had not escaped this misconception. The following footnote was placed there about $BaTiO_3$ which they used to exemplify the structural mechanism of a second-order transition: "To avoid misunderstanding it should be noted that in the particular case of $BaTiO_3$ atomic shifts experience a finite jump, although a small one, so that the transition is still that of first order". A size of the jump is irrelevant: all first-order phase transitions occur by nucleation and growth, rather than by cooperative atomic shifts.

These were examples characteristic of the whole picture. Such inseparable attributes of first-order phase transitions as nucleation, moving interfaces and a temperature range of two-phase coexistence were missing. The first-second-order classification being still recognized *de jure*, was almost abandoned *de facto*. The original intent (definitely shared by Landau) to distinguish the two antipodal types was replaced by blurring all boundaries between them in attempts to regard them as resulted from fluctuations in the bulk. The desire to treat all phase transitions as second order has turned out irresistible. The theoretical physicists wanted to apply their powerful tool - statistical mechanics. Unfortunately, it is applicable only to those solid-state phase transitions that have not yet been found.

### 6. Scaling all solid-state phase transitions

Next theoretical step was the "scaling renormalization group" theory of the 1970's [28,29]. Even though it was a *theory of second-order phase transitions*, this limitation soon vanished in the same way as it happened to the Landau's theory: it became simply a *theory of phase transitions* [30]. In the instances when first-order phase transitions were not ignored, they were incorporated into the new theory. As one author claimed, "the scaling theory of critical phenomena has been successfully extended for classical first order transitions…" [31]. Taken into account the actual physical process illustrated in Fig. 1, such "extension" was a meaningless mathematical exercise.

### 7. Nucleation-and-growth quantum phase transitions?

The ensuing theoretical development was "quantum phase transitions", put forward in the last decade of 20th century [32,33]. This theory considers all solid-state phase transitions being "classical", except their special form, called "quantum", occurring at or close to $0^o$ K. The "classical" phase transitions are claimed to be continuous and fluctuation-based, with "critical points", *etc*. The "quantum" ones are a "critical phenomenon" as well, differing from the "classical" by absence of the thermal fluctuations. A problem with this theory is that "classical" phase transitions are actually nucleation-and-growth. Even Landau with his statement that phase transitions are "usually" first order was set aside when he became an obstacle. There is no reason for the transitions that occur close to $0^o$ K not to be nucleation-and-growth. More detailed analysis of the theory is given in Ref. 21 (2nd Ed., Addendum B).

The incorporation of first-order phase transitions into the theory of "quantum" phase transitions followed: once again, nucleation and crystal growth became a homogeneous process and a "critical phenomenon". Lastly, the theoretical generalization has achieved its culmination when "scaling ideas [were applied] to quantum first order transitions" [31].

### 8. Soft-mode, displacive, topological, *etc.*

To complete the picture, some independent theoretical branches should also be mentioned, all disregarding the real nucleation-and-growth mechanism. They are: *soft-mode* concept, *displacive* phase transitions, and *topological* phase transitions.

The *soft-mode* concept [34-37] claims phase transitions to occur by sudden cooperative "distortion" of the initial phase as soon as one of the low-frequency optical modes "softens" enough toward the transition temperature. Hear we deal with the cooperative macroscopic changes not permitted by thermodynamics (conditions *'A'* and *'B'* in section 1 above). More details on this subject can be found in Ref. 21 (Sec. 1.6, 1.7).

The *displacive* phase transitions were rather an idea then a theory, and no experimental proof of that idea ever existed. They were assumed from comparisons of the initial and final structures when they "looked similar". The idea was put forward by Buerger in the 1950's [26] as deformation/distortion of the original structure by cooperative displacements of the atoms/molecules in the crystal lattice without breaking their chemical bonding. It did not work well, since some bonding still had to be broken. Nevertheless, it is presently sufficient for a phase transition to be called *displacive* if the two crystal phases are "sufficiently similar". If they are not, an imaginary trajectory is constructed to achieve the transformation in several intermediate "displacive" steps. In such a case the phase transition is called *topological*. These two imaginary mechanisms cannot materialize on the same reason: they are *cooperative macroscopic jumps*. Besides, they are not needed, considering that phase transitions can (and do) occur by nucleation and growth. More on these two types are given in Ref. 21 (2nd Ed., Addendum C).

## 9. Search for truly second-order transitions

Landau himself was unable to produce a correct example of structural second-order phase transition, and no one filled the void since. Ascribing a second order to structural phase transitions is still not rare, but it is always superficial, being a side product in the investigation of something else. Probably, not observing hysteresis or a large "jump" in the recording property, or taking the latent heat for heat capacity, was the "criterion". A detailed experimental investigation of a few cases seemingly lacking hysteresis and most reminiscent to be second order [17] revealed that the crystal structure was layered, the hysteresis, even though very small, existed, and the transition proceeding by the interface propagation.

The rotational order-disorder phase transitions are another instructive example. *Orientation-disorder crystals* (ODC) are a mesomorphic state in which the constituent particles are engaged in thermal hindered rotation, while retaining the 3-D translation crystal order. It seemed a common sense to claim that phase transitions CRYSTAL - ODC are of second order. But the hope that second-order phase transitions found at last an ideal area of their application quickly faded. Landau and Lifshitz [6] warned: "There are statements in literature about a connection between second-order phase transitions and emerging rotating molecules in the crystal. This belief is erroneous…" After that it still took years for the problem to become settled. It was investigated in Ref. 21 (Sec. 2.7) and shown that such representative candidates for second-order CRYSTAL - ODC phase transition as $CBr_4$, $C_2Cl_6$, $CH_4$, $NH_4Cl$, $CBr_4$ - are realized by nucleation and growth. In the case of $C_2Cl_6$ the photographic pictures were taken [12] exhibiting growing faceted orientation-disordered crystals in the "normal" non-rotational crystal phase. The "disordering" proceeded by nucleation and crystal growth.

From 1970's some theorists abandoned looking for a good example of structural second-order phase transition and turned to ferromagnetic phase transitions [38]. Vonsovskii [39] stated that the theory of second-order phase transitions provided an "impetus" to studies of magnetic phase transitions. In view of the incessantly shrinking availability of second-order phase transitions, ferromagnetic transitions became the most reliable example of their existence, and first of all, the ferromagnetic phase transition in Fe. In 1965 Belov [40] wrote that ferromagnetic and antiferromagnetic transitions are "concrete examples" of second-order phase transitions. His work was devoted to the investigation of spontaneous magnetization and other properties of Ni in the vicinity of the Curie points. The problem was, however, how to extract these "points" from the experimental data which were always "smeared out" and had "tails" on the temperature scale, even in single crystals. Unfortunately for this and other authors, they were actually dealing with all the effects that accompany first-order nucleation-growth phase transitions, namely, the temperature ranges of phase transitions and related pseudo-anomalies.

Just a few years later it was recognized that some ferromagnetic phase transitions were of the first order. In the book on magnetism by Vonsovskii [39] about 25 such phase transitions were already listed. They were interpreted in the usual narrow-formal manner as those exhibiting "abrupt" changes and / or hysteresis of the magnetization and other properties. A puzzling fact of their existence led to theoretical and experimental studies. It was always assumed that magnetization was the cause of phase transitions, while changes in the crystal parameters, density, heat capacity, etc.- the accompanying effects. The idea that change in the state of magnetization is *caused* by change in the crystal structure has not emerged. The conventional theory was in a predicament: the Curie point was not a point any more, and was rather a range of points and, even worse, was a subject to temperature hysteresis. It was not realized that a first-order phase transition meant nucleation and growth, and not a critical phenomenon. The problem of the first-order ferromagnetic phase transitions had not been resolved.

The thermodynamic theory that treats ferromagnetic phase transitions as being continuous lost its grounds. It cannot be applied even to such basic ferromagnets as Fe, Ni and Co. A "discontinuity" of the Mössbauer effect in the case of Fe was first reported in 1962 by Preston *et al*. [41], and later in more detail by Preston [42], who stated that this "might be interpreted as evidence for a first-order transition". As for Ni, the title "Mössbauer Study of Magnetic First-Order Transition in Nickel" [43] speaks for itself. The *ferromagnetic - paramagnetic* phase transition in Fe was analyzed in Ref. 21 (Sec. 4.2.3, 4.7) and concluded to be a case of nucleation and growth. Finally, the ferromagnetic phase transitions in Fe, Ni and Co were confirmed to be first order by direct experiments [23]. Yet, Fe is still used as the best example of a continuous ferromagnetic phase transition (*e.g*., [33]). Evidently, a better example has not been found. As in case of structural phase transitions, a steady process of second-to-first-order reclassification is going on. The Google search for "first order magnetic transition", taken in January 2011 produced 2,530,000 hits, more by 20% than hits for "second order magnetic transition". Many ferromagnetic phase transitions are presently called "magnetostructural", thus assuming that there are those not being structural. A question why some ferromagnetic transitions are combined with simultaneous structural change, and others are not, is not raised. Simple explanation of that incoherence given in Ref. 21 (Chapter 4) is: *all* ferromagnetic phase transitions resulted from change of crystal structure. It is structural phase transition that brings a magnetization change about, and not the other way round.
.

It is presently widely accepted that "most ferroelectric phase transitions are not of second order but first [44]. "Only very few ferroelectrics…have critical or near critical transitions…the majority having first-order transitions" [45] and materialize by nucleation and growth [46]. And what about the remaining very few? That the phase transition in $BaTiO_3$ was reclassified to first order was mentioned above. The same happened to $KH_2PO_4$, even though "for years this crystal had been regarded as a typical representative of ferroelectrics undergoing second-order phase transition" [47]. The transition in TGS (tri-glycine sulfate) was believed to be the most typical second-order ferroelectric phase transition. As soon as small single-domain TGS samples were used, the characteristics of first-order phase transition were found [48]. Jumps of the electric properties and small (~ 0.2 $^o$C ) hysteresis were detected [47]. The phase transition CUBIC - TETRAGONAL in $SrTiO_3$ at 105$^o$ K was confidently regarded to be second order, but later became a subject of discussion "whether pure $SrTiO_3$ possesses a first order transition or not. This question has not been clarified yet…" [49]. If the correct criteria (heterophase state, hysteresis, *etc.*) were applied to the already accumulated experimental data, its first-order mechanism would become obvious.

As happened in other cases, a second-order nature of *superconducting* phase transitions was initially taken for granted, but later became debatable. Many superconducting phase transitions has been directly named first order. The Google search for "first order superconducting phase transition", taken in January 2011, already produced 242,000 hits, more by 22% than hits for "second order superconducting phase transition". The presently available experimental data, if properly taken into account, would attest that all superconducting phase transitions are first order. They are accompanied by sharp change in physical properties indicative of a discontinuous change of the crystal structure. This should not occur in second-order transitions by their definition.

A well-documented example of "pure" second-order superconducting transition does not exist. The claims about second order are usually based on the absence of latent heat. However, the latent heat can be small and simply avoided detection. More importantly, it has been proven (Ref. 21, Chapter 3) that the utilized calorimetric methods of measurement do not separate *latent heat* from *heat capacity*, ascribing their combined effect to the latter. Detection of an interface, or a two-phase coexistence, or a hysteresis would proof the first-order of all those transitions. These reliable characteristics are frequently present in the experimental data, but their role as indicators of a first order not always recognized. Some superconducting phase transitions are called "weakly first order" to treat them as second order. However, first-order phase transitions, "weakly" or not, are a local "molecule-by-molecule" process.

First-order superconducting phase transitions should have serious implications for the theories of superconductivity involving mechanism of the phase transition. The point is that all first-order phase transitions, including superconducting, are a nucleation-and-growth structural rearrangement. While comparison of the initial and resultant crystal structures may be useful, or even vital for understanding of the nature of superconductivity, the process of their crystal rearrangement is hardly specific to this kind of phase transitions.

### 10. Why they are not found

The conditions *'A'* and *'B'* cited in the beginning are *necessary* ones. They represent two different hypothetical ways to rich the final state. The way satisfying the condition *'A'* is the actually realized nucleation-and growth mechanism. Clearly, it is *sufficient*. The condition *'B'* would be *sufficient* if the human-proposed cooperative mechanism could in some cases successfully compete with the nucleation-and-growth. Then theoretical physicists would have an area for application of their talents, their knowledge of statistical mechanics and their belief in the fluctuation power in everything. But comparative analyses of the energy required by the two mechanisms have not been done. The important questions like *why* the phase transition in $BaTiO_3$ is of first order, while in $SrTiO_3$ of second order, were not raised. Reliable examples of second-order phase transitions have not been found. Yet, second-order type of phase transitions was assumed to be a reality.

But Nature had its own agenda, namely, to make its natural processes (a) universal, (b) simple and (c) the most energy-efficient. It produced a better process than the most brilliant human beings, even Nobel Prize winners, could invent. Solid-state phase transition by nucleation and growth, as described above in section 5, is that process. It is more universal, simple and energy-efficient than critical-dynamic theories offered. It is universal because it is just a particular manifestation of the general crystal growth in liquids and solids; even magnetization by magnetic field is realized by nucleation and growth [50]. It is also as simple as crystal growth. It is most energy-efficient because it needs energy to relocate one molecule at a time, and not the myriads of molecules at a time as a cooperative process requires. This is why true second-order phase transitions will never be found. The first-second-order classification should be laid to rest.

**References**

[1]   M. Azbel, Preface to R. Brout, *Phase Transitions* (Russian ed.), Mir, Moscow (1967).
[2]   P. Ehrenfest, *Leiden Comm. Suppl.*, No. 75b (1933).
[3]   E. Justi and M. von Laue, *Physik Z.* **35**, 945; *Z. Tech. Physik* **15**, 521 (1934).
[4]   L. Landau, in *Collected Papers of L.D. Landau*, Gordon & Breach (1967), p.193. [*Phys. Z. Sowjet.* **11**, 26 (1937); **11**. 545 (1937)].
[5]   L. Landau and E. Lifshitz, in *Collected Papers of L.D. Landau*, Gordon & Breach (1967), p.101. [*Phys. Z. Sowiet* **8**, 113 (1935)].
[6]   L.D. Landau and E.M. Lifshitz, *Statistical Physics*, Addison-Wesley (1969).
[7]   Y. Mnyukh, *J. Phys. Chem. Solids*, **24** (1963) 631.
[8]   A.I. Kitaigorodskii, Y. Mnyukh, Y. Asadov, *Soviet Physics - Doclady* **8** (1963) 127.
[9]   A.I. Kitaigorodskii, Y. Mnyukh, Y. Asadov, *J. Phys. Chem. Solids* **26** (1965) 463.
[10]  Y. Mnyukh, N.N. Petropavlov, A.I. Kitaigorodskii, *Soviet Physics - Doclady* **11** (1966) 4.
[11]  Y. Mnyukh, N.I. Musaev, A.I. Kitaigorodskii, *ibid.* **12** (1967) 409.
[12]  Y. Mnyukh, N.I. Musaev, *ibid.* **13** (1969) 630.
[13]  Y. Mnyukh, *ibid.* **16** (1972) 977.
[14]  Y. Mnyukh, N.N. Petropavlov, *J. Phys. Chem. Solids* **33** (1972) 2079.
[15]  Y. Mnyukh, N.A. Panfilova, *ibid.* **34** (1973) 159.
[16]  Y. Mnyukh, N.A. Panfilova, *Soviet Physics - Doclady* **20** (1975) 344.
[17]  Y. Mnyukh *et al.*, *J. Phys. Chem. Solids* **36** (1975) 127.
[18]  Y. Mnyukh, *J Crystal Growth* **32** (1976) 371.
[19]  Y. Mnyukh, *Mol. Cryst. Liq. Cryst.* **52** (1979) 163.
[20]  Y. Mnyukh, *ibid.*, **52** (1979) 201.
[21]  Y. Mnyukh, *Fundamentals of Solid-State Phase Transitions, Ferromagnetism and Ferroelectricity*, Authorhouse, 2001 [or 2$^{nd}$ (2010) Edition].
[22]   Ref, 21, Sec.4.2.3 and Fig.4.2.
[23]  Sen Yang *et al.*, *Phis. Rev.* **B 78**,174427 (2008).
[24]   A.D. Bruce and R.A. Cowley, *Structural Phase Transitions*, Taylor and Francis (1981).
[25]   J.C. Toledano and P. Toledano, *The Landau Theory of Phase Transitions*, World Sci. (1986).
[26]   M.J. Buerger, *Kristallografiya*, **16**, 1048 (1971) [*Soviet Physics - Crystallography* **16**, 959 (1971)].
[27]  D.R. Moore et al., *Phys. Rev.* **B 27**, 7676 (1983).
[28]  K.G. Wilson, *Phys. Rev.* B **4** (1971), 3174.
[29]  K.G. Wilson,. *Scientific American*, August 1979.
[30]  M.L.A. Stile: Press Release: *The 1982 Nobel Prize in Physics*, http://nobelprize.org/nobel_prizes/physics/laureates/1982/press.html.
[31]  M. A. Continentino, cond-mat/0403274.
[32]  S. Sachdev, *Quantum Phase Transitions*, Cambridge University Press (1999).
[33]  M. Vojta, cond-mat/0309604.
[34]  W. Cochran, *Adv. Phys..* **9**, 387 (1960).
[35]  *Structural Phase Transitions and Soft Modes*, Ed. E.J. Samuelson and J. Feder., Universitetsfurlaget, Norway (1971).
[36]  J. F. Scott, *Rev. Mod. Phys.* **46**, 83 (1974).
[37]  G. Shirane, *Rev. Mod. Phys.* **46**, 437 (1974).

[38] H.E. Stanley, *Introduction to Phase Transitions and Critical Phenomena*, Clarendon Press (1987).
[39] S.V. Vonsovskii, *Magnetism*, vol. 1 & 2, Wiley (1974).
[40] K.P. Belov, *Magnetic Transitions*, Boston Tech. Publ. (1965).
[41] R.S. Preston, S.S. Hanna and J. Heberle, *Phys. Rev*. **128**, 2207 (1962).
[42] R.S. Preston, *Phys.Rev.Let.*, **19**, 75 (1967).
[43] A.A. Hirsch, *J.Magn.Magn.Mater*. **24**, 132 (1981).
[44] M.E. Lines and A.M. Glass. *Principles and Applications of Ferroelectrics and Related Materials*, Clarendon Press (1977).
[45] N.G. Parsonage and L.A.K. Staveley, *Disorder in Crystals*, Clarendon Press (1978).
[46] *e.g*., V.M. Ishchuk, V.L. Sobolev, *J. Appl. Phys.* **92** (2002) 2086.
[47] I.S. Zheludev, *The Principles of Ferroelectricity*, Atomizdat, Moscow (1973, Rus.).
[48] G.G. Leonidova, *Docl. Acad. Nauk SSSR* **196**, 335 (1971).
[49] J.O. Fossum *et al.*, *Solid State Comm.* **51** (1984), 839.
[50] Y. Mnyukh, http://arxiv.org/abs/1101.1249.